\documentclass[prl,twocolumn,showpacs,preprintnumbers,amsmath,amssymb,superscriptaddress,floatfix]{revtex4}

\usepackage{graphicx}
\usepackage{dcolumn}
\usepackage{bm}
\usepackage{epsf}

\begin{document}


\title{Evidence for the constancy of \textbf{\textit{U}} in the Mott transition of V$_{2}$O$_{3}$}


\author{H. Fujiwara}
\affiliation{Graduate School of Engineering Science, Osaka University, Toyonaka, Osaka 560-8531, Japan}
\affiliation{II. Physikalisches Institut, Universit{\"a}t zu K{\"o}ln, Z{\"u}lpicher Stra{\ss}e 77, 50937 K{\"o}ln, Germany}
\author{A. Sekiyama} 
\affiliation{Graduate School of Engineering Science, Osaka University, Toyonaka, Osaka 560-8531, Japan}
\affiliation{SPring-8/RIKEN, Kouto 1-1-1, Sayo, Hyogo 679-5198, Japan}
\author{S.-K. Mo}
\affiliation{Randall Laboratory of Physics, University of Michigan, Ann Arbor, Michigan 48109, USA}
\affiliation{Department of Physics, Stanford University, Stanford, California 94305, USA}
\author{J. W. Allen}
\affiliation{Randall Laboratory of Physics, University of Michigan, Ann Arbor, Michigan 48109, USA}
\author{J. Yamaguchi}
\author{G. Funabashi} 
\affiliation{Graduate School of Engineering Science, Osaka University, Toyonaka, Osaka 560-8531, Japan}
\author{S. Imada}
\affiliation{Graduate School of Engineering Science, Osaka University, Toyonaka, Osaka 560-8531, Japan}
\author{P. Metcalf}
\affiliation{Department of Physics, Purdue University, West Lafayette, Indiana 47907, USA}
\author{A. Higashiya}
\affiliation{SPring-8/RIKEN, Kouto 1-1-1, Sayo, Hyogo 679-5198, Japan}
\author{M. Yabashi}
\affiliation{SPring-8/RIKEN, Kouto 1-1-1, Sayo, Hyogo 679-5198, Japan}
\affiliation{SPring-8/JASRI, Kouto 1-1-1, Sayo, Hyogo 679-5198, Japan}
\author{K. Tamasaku}
\affiliation{SPring-8/RIKEN, Kouto 1-1-1, Sayo, Hyogo 679-5198, Japan}
\author{T. Ishikawa}
\affiliation{SPring-8/RIKEN, Kouto 1-1-1, Sayo, Hyogo 679-5198, Japan}
%
\author{S. Suga} 
\affiliation{Graduate School of Engineering Science, Osaka University, Toyonaka, Osaka 560-8531, Japan}
\affiliation{SPring-8/RIKEN, Kouto 1-1-1, Sayo, Hyogo 679-5198, Japan}

\date{\today}

\begin{abstract}
We have performed high-resolution hard X-ray photoemission spectroscopy for the metal-insulator transition (MIT) system (V$_{1-x}$Cr$_x$)$_{2}$O$_{3}$ in the paramagnetic metal, paramagnetic insulator and antiferromagentic insulator phases.  The quality of the spectra enables us to conclude that the on-site Coulomb energy $U$ does not change through the MIT, which eliminate all but one theoretical MIT scenario in this paradigm material. 
\end{abstract}


\maketitle
%

\begin{figure}
\begin{center}
\includegraphics[width=8cm,clip]{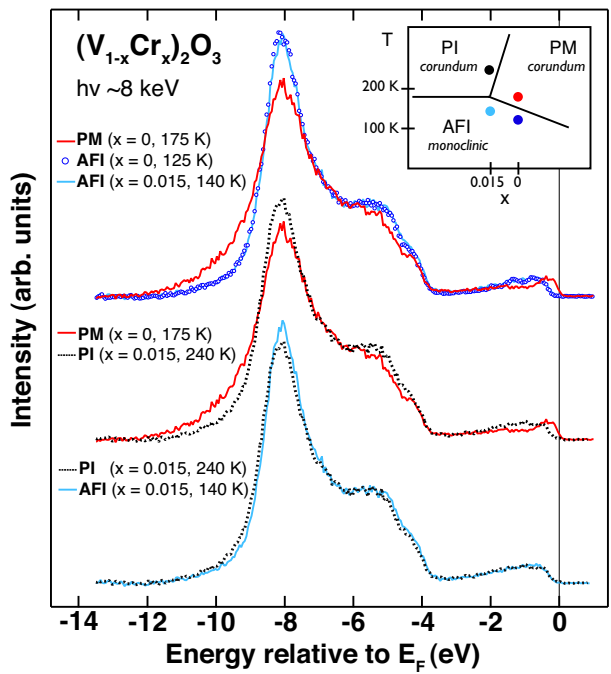}
\caption 
{(color online). Entire valence band spectra of (V$_{1-x}$Cr$_{x}$)$_{2}$O$_{3}$ ($x$ = 0 and 0.015) in the PM and AFI phases (top), those in the PM and PI phases (middle), and those in the PI and AFI phases (bottom). Inset shows the schematic phase diagram of (V$_{1-x}$Cr$_{x}$)$_{2}$O$_{3}$.}
\label{Val_wide_narrow}
\end{center}
\end{figure}

The Mott-Hubbard metal-insulator transition (MH-MIT) is a fundamental phenomenon not only for strongly correlated physics~\cite{Imada_MIT1998} but also for solid state physics generally.  Dynamic mean field theory (DMFT) ~\cite {DMFT_prototype} has provided a new conceptual framework for describing how the MH-MIT occurs in the one-band Hubbard model when the on-site Coulomb repulsion $U$ exceeds a critical value $U_{c}$ relative to the inter-site hopping energy $W$.  Nonetheless, even when DMFT is combined with realistic electronic structure calculations in the local density approximation (LDA +DMFT), an internally consistent description of the MIT $\emph{as it is observed in nature}$ has not yet emerged.  This is true even for the most heavily studied paradigmatic material V$_2$O$_3$.  This material shows a first order transition from a high-temperature paramagnetic metal (PM) phase to a low-temperature antiferromagnetic insulator (AFI) phase at $T_{\textrm{N}}\approx155$ K, accompanied by a structural change from the corundum phase to the monoclinic phase~\cite{McWhan,Dernier1970,Moon}. Slight Cr-substitution on the V sites develops a paramagnetic insulator (PI) phase with the same corundum structure above $T_{\textrm{N}}\approx180$ K as shown in the inset of Fig.~\ref{Val_wide_narrow}.  In the comparison between early LDA+DMFT calculations\cite{DMFT_LDA_Held2001,Mo2003,Mo2006,KellerHeld2004} and valence band photoemission spectra, the $U$ value consistent with spectroscopy~\cite{spectral weight ratio} is too small to allow the MIT for the $W$ values found in the LDA calculations for the various phases.

This basic tension concerning $U$ has proved to be continuing and pervasive even though the theory has been made more sophisticated.  Since changes in the V $3d$ orbital occupation are reported through all these transitions~\cite{Park2000}, an interplay among the spin-, charge- and orbital-degrees of freedom is thought to be essential for MIT.  More recent LDA + DMFT studies suggest that an effective trigonal crystal field splitting leads to a redistribution of the orbital populations (known as the orbital selective MIT)~\cite{Laad2006,Poteryaev2007}.  Nonetheless $U$ must still be changed through the transition in the cited theories.  A very interesting alternative concept is that the orbital polarization induced by a slight enhancement of the effective crystal field splitting can reduce $U_{c}/W$ and thereby facilitate the MIT~\cite{Pavarini2004, Poteryaev2008} even though $U$ is unchanged.  Thus there are now only two possibilities on the table for V$_2$O$_3$, either $U$ or $U_{c}$ changes in the MIT.  A direct experimental determination of the differences in the bulk electronic structures and the value of $U$ in all three phases is essential to move the issue forward.

In this Letter, we report a state of the art hard X-ray photoemission spectroscopy (HAXPES) study to tackle this problem. Since $U/W$ is known to be much different for the surface and the bulk, bulk sensitive HAXPES with $h\nu=5-8$ keV~\cite{Kamakura2004,Taguchi2005,suga2005,Panaccione2006,Yamasaki2007} is essential by virtue of its large probing depth ($>50$ {\AA} at $h\nu\sim8$ keV). The very high quality of our HAXPES spectra relative to that of earlier studies enables a detailed analysis to identify in all three phases the incoherent part of the V $3d$ spectrum, i.e. the lower Hubbard band (LHB) that defines $U$ on the photoemission side of the Fermi level $E_{\textrm{F}}$.  Thereby we obtain very strong evidence that $U$ stays essentially constant through the MIT in (V$_{1-x}$Cr$_x$)$_{2}$O$_{3}$, leaving a change in $U_{c}/W$ as the viable scenario for the MIT.

HAXPES was performed at BL19LXU in SPring-8 with use of an MBS A1-HE hemispherical analyzer system. The linearly polarized light at $\sim8$ keV was delivered from an in-vacuum 27-m long undulator~\cite{Yabashi2001}. The beam was focused onto the sample within $50\times100$ ${\mu}$m$^2$. The overall energy resolution was set to 220 meV for the wide scan of the valence band and 130 meV for the high-resolution mode near $E_{\textrm{F}}$ as confirmed by the Au Fermi-edge. Well-annealed oriented single crystalline samples were cleaved \textit{in situ} in a vacuum of $8\times10^{-8}$ Pa. The measurement was performed above and below the temperatures of the PM-AFI and PI-AFI transitions (inset of Fig.~\ref{Val_wide_narrow}). Soft X-ray photoemission spectroscopy (SXPES) measurements were carried out at BL25SU in SPring-8 with a comparable energy resolution~\cite{Mo2003,Mo2006}.

Figure~\ref{Val_wide_narrow} shows the HAXPES spectra of (V$_{1-x}$Cr$_{x}$)$_{2}$O$_{3}$ ($x$ = 0 and 0.015) of the entire valence band. A Shirley-type inelastic integral background is subtracted from each spectrum and the spectra are normalized by their integrated area over the whole energy range of the valence band in all phases. In Fig~\ref{Val_wide_narrow}, one can recognize that the AFI spectra of the samples with and without Cr-doping are very similar to each other, demonstrating the negligible effect of disorder on the electronic structure. 
All phases show a large peak structure around $-$8 eV, which is derived from O $2p$ states hybridized with V $4s$ and $3d$ states~\cite{Woicik2007,Koethe-LHT2009,Sekiyama_daiamond,Papalazarou_v2o3crosssection}. The width of the $-$8 eV peak of the PM phase is wider than those in the AFI and PI phases. 

\begin{figure}
\begin{center}
\includegraphics[width=8cm,clip]{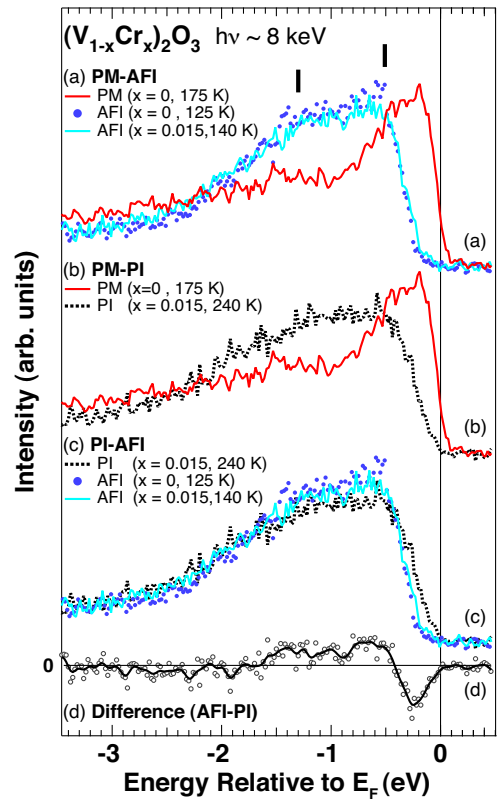}
\caption 
{(color online). High-resolution HAXPES spectra of (V$_{1-x}$Cr$_{x}$)$_{2}$O$_{3}$ ($x$ = 0 and 0.015) near $E_{\textrm{F}}$ in the PM and AFI phases (a), those in the PM and PI phases (b), and those in the PI and AFI phases (c). (d) The difference between the AFI and PI spectra (open circle), which is estimated by averaging the two results for the $x$ = 0 and 0.015 AFI spectra. The thick solid line is the smoothed difference.}
\label{SpecWeightTrans}
\end{center}
\end{figure}

Figure~\ref{SpecWeightTrans} displays the high-resolution HAXPES spectra reflecting the bulk V $3d$ electronic states. The background is subtracted in the same manner as for Fig.~\ref{Val_wide_narrow}. It is useful to first compare the PM and PI spectra (Fig.~\ref{SpecWeightTrans}(b)), for which the problem is simplified by excluding the roles of both the magnetic long-range ordering and the structural phase transition. The PM spectrum in Fig.~\ref{SpecWeightTrans}(b) shows the prominent quasiparticle (QP) peak just below $E_{\textrm{F}}$. In addition, the small bump structure observed around $-1.3$ eV is ascribed to the incoherent part corresponding to the LHB~\cite{Mo2003,Mo2006,Panaccione2006}. In contrast the PI spectrum shows a gap opening, with a strong spectral weight transfer across the PM-PI transition. 

A strong spectral weight transfer is also observed across the PM-AFI transition as shown in the Fig~\ref{SpecWeightTrans}(a), indicating that the crystal symmetry change in this transition is not a major factor of the MIT in (V$_{1-x}$Cr$_{x}$)$_{2}$O$_{3}$.  There are however noticeable differnces in the AFI and PI lineshapes.  First the AFI phase band gap is larger(220 meV from $E_{\textrm{F}}$) and the threshold is noticeably sharper than for the PI phase, in agreement with the results of optical measurements~\cite{optcond} and the previous SXPES~\cite{Mo2006}. Second the spectra in both the doped and undoped AFI phases do not show a single LHB peak as predicted by the one-band Mott-Hubbard model, but consist of multi-components as indicated by the two vertical bars for the shoulder and peak structures at around $-1.3$ and $-0.5$ eV, respectively. These differences are seen directly in the spectra for the PI-AFI transition shown in Fig.~\ref{SpecWeightTrans}(c) and also in the negative and positive peaks in the regions near $-$0.25 and $-0.5\sim-0.75$ eV in the difference spectrum in Fig.~\ref{SpecWeightTrans}(d).  The AFI structure at $-1.3$ eV can most likely be ascribed to the LHB. On the other hand, the $-0.5$ eV peak was not predicted by the early LDA+DMFT calculations for V$_{2}$O$_{3}$~\cite{DMFT_LDA_Held2001}. We notice in Fig.~\ref{SpecWeightTrans}(c) that the tails on the lower energy side (from $-$2 to $-$3 eV) are almost identical, suggesting that $U$ (as observed on the PES side of $E_{\textrm{F}}$) may be essentially the same in these two phases. In order to confirm this conclusion, it is essential to know the energy positions of the LHB in all phases.  To do that we must first establish firmly the origin of the -0.5 eV peak in the AFI phase.

\begin{figure}
\begin{center}
\includegraphics[width=8cm,clip]{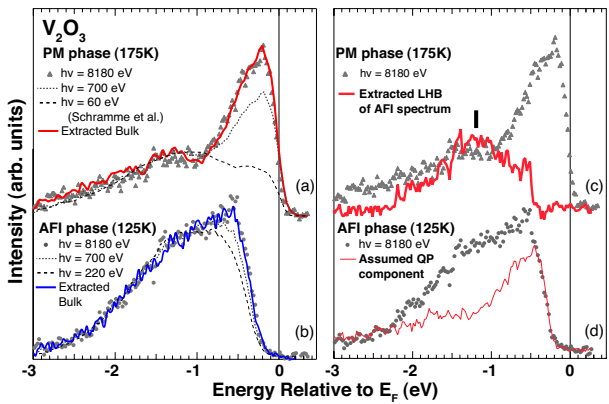}
\caption 
{(color online). (left panel) Photon energy dependence of the V $3d$ spectral weight near $E_{\textrm{F}}$ for V$_{2}$O$_{3}$ taken by HAXPES and SXPES in the PM phase (a) and in the AFI phase (b). Solid lines in (a) and (b) show the bulk component extracted from the PES spectra taken at 700 and 60 eV~\cite{Schramme2000} (700 and 220 eV) photons in the PM (AFI) phase, respectively. (right panel) Comparison between the PM spectrum and the extracted LHB of AFI spectrum (c). The extracted LHB of AFI phase in (c) is evaluated by the procedure mentioned in the text in detail.} 
\label{HAXSX}
\end{center}
\end{figure}

In Fig.~\ref{HAXSX}(a and b) we show the $h\nu$ dependences of the spectral weight near $E_{\textrm{F}}$ for pure V$_{2}$O$_{3}$ for both the PM and AFI phases. For easy comparison of the intensity near $E_{\textrm{F}}$, the background is here subtracted in the same manner as in refs.~\cite{Mo2003,Mo2006} and the spectra are normalized below $-1$ eV. As shown in Fig.~\ref{HAXSX}(a) for the PM phase, the relative weight between $E_{\textrm{F}}$ and $-1$ eV representing the QP-peak increases remarkably with increasing $h\nu$ from 60 eV~\cite{Schramme2000} to 8180 eV in the PM phase. This development of the QP-peak is not due to a change of the relative cross-sections of V $3d$ and O $2p$ states, but to the enhancement of the bulk contribution in accordance with the increase of the probing depth~\cite{Mo2003,Mo2006}. We also show that the extracted bulk component from the photoemission spectra at $h\nu=700$ and 60 eV~\cite{Mo2003,BulkSurfSep_Method} is in full agreement with the HAXPES spectrum. Thus the increase of the relative intensity of the QP-peak can be interpreted as due to the reduction of the $U/W$ in the bulk caused by the wider bandwidth $W$ in the bulk than in the surface. We note in passing that the finding of a larger QP peak in the HAXPES spectrum demands a smaller value of $U$, which exacerbates the problem that $U$ is then too small to enable the MIT.

The $-0.5$ eV peak in the AFI phase also increases with increasing $h\nu$, as seen very clearly in Fig.~\ref{HAXSX}(b). Although the presence of this spectral feature has been inferred in previous studies~\cite{Mo2006}, it is here more clearly seen by comparison of SXPES at 700 and 220 eV and directly detected owing to the higher bulk sensitivity of HAXPES. The development of the $-$0.5 eV structure in the AFI phase with increasing bulk sensitivity is thus in parallel with that of the QP-peak in the PM phase. It is significant that such a peak at $-$0.5 eV is not seen in the PI phase as $hv$ is increased.

Following the early suggestion by Rozenberg \textit{et al.}~\cite{Rozenberg1996}, the leading edge and peak in the AFI spectrum has been ascribed~\cite{Sangiovanni2006} to a QP induced by long-range spin coherence. The clear observation of the $-0.5$ eV peak and particulary its $h\nu$ dependence in the present AFI phase data provide needed direct evidence that this QP assignment is indeed correct. The AFI QP spectral weight depends on $U/W$ much like that of the QP-peak in the metallic phase~\cite{Sangiovanni2006}. Following the discussion of Rozenberg \textit{et al.}~\cite{Rozenberg1996}, the weight scales with the magnetic exchange $J{\sim}t{^2}/U$; $t$ is the hopping integral, which is related to $W$ as $W{\propto}2t$. Since $U/W$ becomes smaller (namely, $t$ is larger) in the bulk, the bulk gives the larger $J$ inducing the larger QP-peak in the AFI spectrum. Thus the similar $h\nu$ dependences of the QP-peaks of the AFI and PM phases can be consistently understood. Finally, DMFT also predicts~\cite{Rozenberg1996, Sangiovanni2006} that the gap size of the PI phase is smaller than that in the AFI phase because of the magnetic disorder. This is fully consistent with our experimental finding in Fig.~\ref{SpecWeightTrans}(c). The slight transfer of the spectral weight across the PI-AFI transition can now be unambiguously interpreted as due to the development of magnetic long-range ordering, nicely consistent with neutron scattering results~\cite{bao}. 

We now proceed to identify the LHB in the AFI spectrum. For this we have adopted as a working hypothesis that for the resolution of the experiment the QP component in the AFI phase has essentially the same spectral shape as in the PM phase. We shift the PM spectrum and normalize it to realize a good fit to the leading edge of the AFI spectrum as shown by the thin solid curve in Fig.~\ref{HAXSX}(d). Then this spectral shape is subtracted from the AFI spectrum, yielding the LHB in the AFI phase as shown by the thick solid curve in Fig.~\ref{HAXSX}(c). The peak position of the LHB in the AFI phase indicated by the vertical bar is thus estimated as $-1.2\pm0.15$ eV, which is almost degenerate with that of the PM phase. Even when the LHB in the PM is not subtracted from the AFI spectrum, no essential change is seen for the LHB spectral weight in the AFI phase. The possibility that the LHB in the AFI phase is deeper than that in PM phase is definitely excluded. $U$ is defined by the separation of the LHB and the upper Hubbard band (UHB) on the unoccupied side of $E_{\textrm{F}}$, which we do not observe.  However, if $U$ increases in the insulating phases due to a reduction in screening, it is highly unlikely that the UHB can shift while the LHB remains fixed. Therefore we conclude that $U$ stays essentially constant through the MIT in (V$_{1-x}$Cr$_{x}$)$_{2}$O$_{3}$.  

Our result strongly supports the scenario~\cite{Pavarini2004, Poteryaev2008} of the orbital selective Mott-transition in which $U_{c}/W$ changes through the MIT due to the enhancement of the effective trigonal crystal field splitting.  We remark that this idea is also supported by the observation in an optical study~\cite{Allen1971-1973} long ago of uniaxial stress-induced spin flop in Cr$_2$O$_3$.  Considering the stress dependences of Cr$^{3+}$ states in ruby~\cite{Sturge}, the spin flop mechanism could be traced to a sign change in the magnetocrystalline anisotropy, driven primarily by a change in the trigonal crystal field. One infers that the trigonal splitting is very strain sensitive in this crystal structure, which may well be the key to making a unified explanation of the MIT in (V$_{1-x}$Cr$_{x}$)$_{2}$O$_{3}$, i.e. a fixed value of $U$ that is consistent with the PM phase QP weight and with the changing LDA values of $W$ enabling the MIT via the changing $U_{c}/W$.

To summarize, new high resolution HAXPES gives noticeable spectral weight transfers for the bulk electronic states in (V$_{1-x}$Cr$_{x}$)$_{2}$O$_{3}$ ($x$ = 0, 0.015) across all three phase transitions. The QP-peak of the AFI phase due to the long-range magnetic ordering is clearly observed and enables an identification of LHB. The essentially degenerate LHB for all three phases yields strong evidence that $U$ stays constant through the MIT. This finding renders a changing $U_{c}/W$ in an orbital selective Mott-transition as currently the only viable scenario for the MIT in (V$_{1-x}$Cr$_{x}$)$_{2}$O$_{3}$.  An understanding of the MIT as it is observed in nature for a paradigm material may finally be at hand.

We thank L. H. Tjeng, I. Nekrasov, K. Haule, and G. Kotliar for fruitful discussions. 
We express appreciation to S. Komori, M. Obara, Y. Nakatsu, Y. Tomida and M. Y. Kimura for supporting the measurements. This work was supported in part by a Grant-in-Aid for 21st century COE (G18), Global COE (G10), and Scientific Research (18104007, 18684015, 21740229, and 21340101) from MEXT and JSPS, Japan. Work at UM was supported by the U.S. DOE under Contract No. DE-FG02-07ER46379. SKM is funded by US NSF. HF thanks the Alexander von Humboldt Foundation for their support.
%



\begin{thebibliography}{00}
\bibitem{spectral weight ratio} Of greatest quantitative significance are the relative weights of the two V $3d$ spectral features discussed in this paper, the quasiparticle peak and the incoherent part.
\end{thebibliography}
\end{document}